\def\ARXIV{}
\documentclass{article}

\newif\ifdeanonymized
\deanonymizedtrue
\ifdefined\ANONYMOUS\deanonymizedfalse\fi

\ifdefined\ARXIV
  \usepackage[preprint]{neurips_2026}
\else\ifdefined\ANONYMOUS
  \usepackage{neurips_2026}
\else
  \usepackage[main, final]{neurips_2026}
\fi\fi

\usepackage[utf8]{inputenc}
\usepackage[T1]{fontenc}
\usepackage{hyperref}
\usepackage{url}
\usepackage{booktabs}
\usepackage{amsfonts}
\usepackage{amssymb}
\usepackage{amsmath}
\usepackage{nicefrac}
\usepackage{microtype}
\usepackage{xcolor}
\usepackage{graphicx}
\usepackage{multirow}
\usepackage{algorithm}
\usepackage{algorithmic}
\usepackage{enumitem}
\usepackage{placeins}
\usepackage{wrapfig}
\usepackage{tikz}
\usetikzlibrary{arrows.meta}

\setlist[itemize]{leftmargin=*, topsep=2pt, itemsep=1pt, parsep=0pt, partopsep=0pt}
\setlist[enumerate]{leftmargin=*, topsep=2pt, itemsep=1pt, parsep=0pt, partopsep=0pt}
\title{\texorpdfstring{\raisebox{-0.22em}{\includegraphics[height=1.2em]{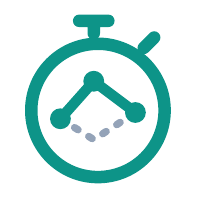}}\hspace{0.22em}}{}LAMaS: A Modular Platform for Learning\\
Low-Latency Multi-Agent Orchestration}

\ifdeanonymized
  \author{%
    Xi Shi \qquad Mengxin Zheng \qquad Qian Lou \\
    LoopLeap Lab, University of Central Florida \\
    \texttt{\{xi320101, mengxin.zheng, qian.lou\}@ucf.edu} \\
    {\normalfont\small GitHub: \url{https://github.com/loopleap-lab/LAMaS}}
  }
\else
  \author{%
    Anonymous Authors \\
    Anonymous Institution \\
    \texttt{anonymous@example.com}
  }
\fi

\begin{document}
\maketitle
\begin{figure}[H]
  \centering
  \includegraphics[width=\linewidth]{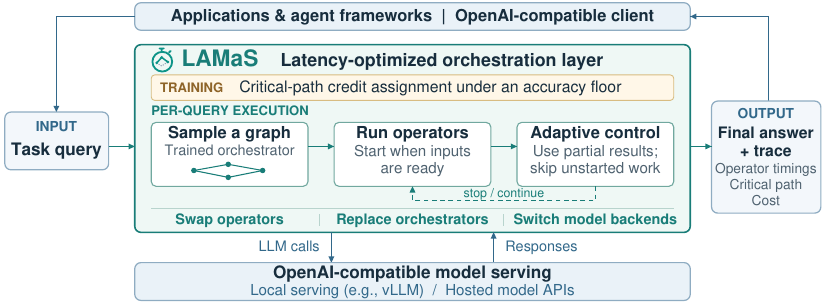}
  \caption{LAMaS connects agent frameworks to model servers and returns an answer and execution trace for each query. We integrate the full pipeline into GDesigner and AgentDropout, retaining each framework's graph builder and retraining the execution controller on its traces.}
  \label{fig:platform}
\end{figure}
\begin{abstract}
In a multi-agent system (MAS), removing an expensive agent call can save tokens without making the answer arrive sooner: latency depends on the longest chain of dependent calls. We present \textbf{L}atency-\textbf{A}ware \textbf{M}ulti-\textbf{a}gent \textbf{S}ystem (\textbf{LAMaS}), a modular platform for learning low-latency orchestration. A framework can send a query to LAMaS and receive an answer and execution trace, or keep its own graph builder and use the platform's latency-optimized orchestration layer. During training, the layer gives operators on the critical path the greatest latency penalty while targeting a floor on expected accuracy. During inference, a lightweight controller uses partial results to decide whether to skip work that has yet to start. Across four benchmarks, LAMaS reduces end-to-end latency by over 50\% relative to the evaluated learning-based MAS baselines while maintaining competitive or better accuracy. GDesigner and AgentDropout retain their graph builders and train the controller on their own traces; both achieve lower latency across the same benchmarks.

\end{abstract}

\section{Introduction}

Multi-agent systems (MAS) use specialized agents to extend the reasoning capabilities of large language models (LLMs) \citep{li2023camel, wu2024autogen, hong2023metagpt}. As the number of agents and interactions grows, inference latency becomes a bottleneck \citep{chen2025kairos, shi2026kvsharearenakvcachereusecontexts, hwrouter2026, kabir2026inframindinfrastructureawaremultiagentorchestration}: tasks in computer use \citep{abhyankar2025osworld} and autonomous research \citep{kang2025win} can take 10--60 minutes. Quantization can speed up individual model calls~\citep{Lou2020AutoQ:, ijcai2021p472}, but the time a user waits also depends on which calls must finish before others can start.

We introduce \textbf{L}atency-\textbf{A}ware \textbf{M}ulti-\textbf{a}gent \textbf{S}ystem (\textbf{LAMaS}), a modular platform that separates an agent framework's choice of operators from how it learns and executes low-latency workflows (Figure~\ref{fig:platform}). Applications can submit queries through an OpenAI-compatible endpoint. Framework developers can instead keep their graph builder and adopt the orchestration layer's training objective and execution controller (Section~\ref{sec:platform}).

\begin{figure}[t]
    \centering
    \includegraphics[width=\textwidth]{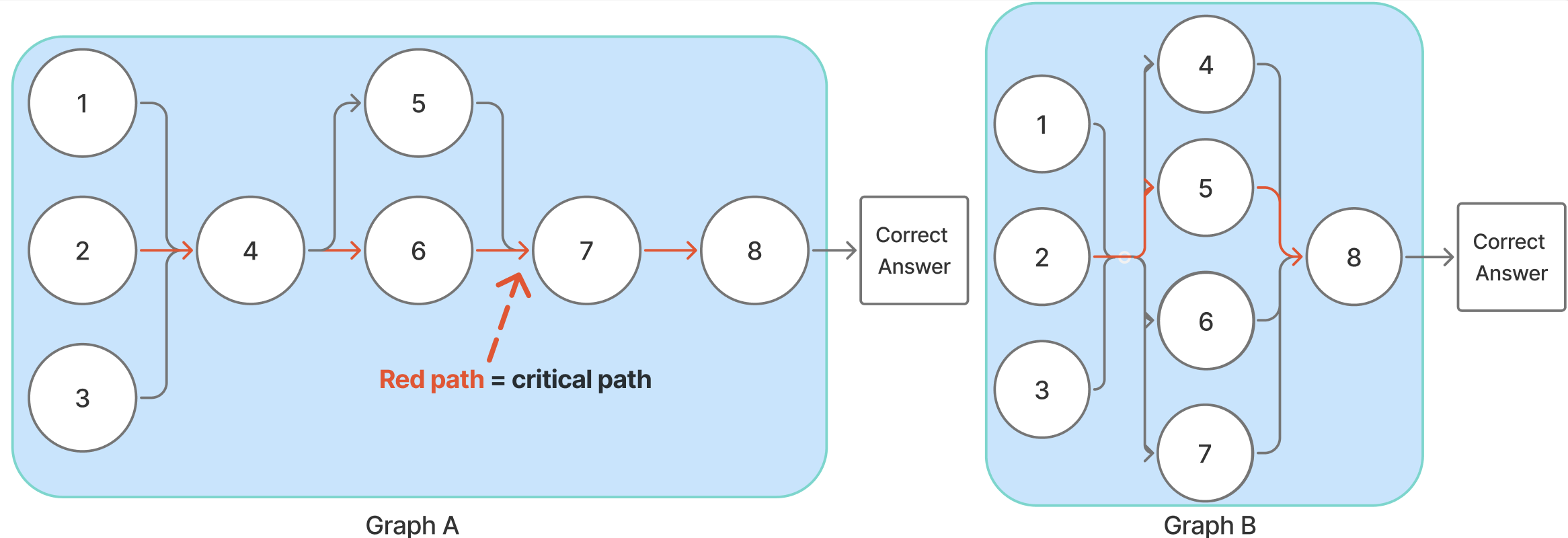}
    \caption{Two execution graphs that produce the same answer at similar cost but differ in latency. Nodes denote agent invocations and arrows denote data flow; red edges mark the critical execution path, the longest chain of dependencies that must complete before an answer is available. Graph~B has a shorter critical path and therefore lower wall-clock latency than Graph~A.}
    \label{fig:motivation}
\end{figure}

Existing methods learn to orchestrate agents for accuracy and inference cost \citep{zhang2025multiagent, wang2025agentdropout, zhang2024g, zhang2025aflow, wang2025evoagentx, wang2025anymac, yao2025can}. Cost adds up across calls; latency follows the \emph{critical path}, the longest chain of dependent calls measured in time. Figure~\ref{fig:motivation} shows why the distinction matters: two graphs produce the same answer at similar cost, but the one with the shorter critical path finishes sooner.

Training and execution call for different decisions. When learning a graph, a uniform latency penalty gives the same signal to operators that delay the answer and those that finish in parallel. LAMaS assigns credit according to each operator's contribution to the critical path and uses Lagrangian relaxation to optimize latency and cost subject to a floor on expected accuracy. At inference time, the graph alone cannot anticipate agreement among early answers. As operators finish, a lightweight controller scores the partial execution under the same objective and decides whether to start more work.

We evaluate LAMaS on GSM8K \citep{cobbe2021training}, HumanEval \citep{chen2021evaluating}, MATH \citep{hendrycks2021measuring}, and MMLU-Pro \citep{wang2024mmlu} against MaAS \citep{zhang2025multiagent}, GDesigner \citep{zhang2024g}, and AgentDropout \citep{wang2025agentdropout}. LAMaS has the lowest latency of the evaluated MAS on all four benchmarks, with competitive or better accuracy. We also integrate the layer into GDesigner and AgentDropout: each retains its own graph builder, trains with critical-path credit, and fits the controller to its own execution traces. Both integrations reduce latency across all four benchmarks (Table~\ref{tab:generalizability}).

\section{Related Work}

\paragraph{LLM-Based Multi-Agent Systems.} Multi-Agent Systems built on LLMs decompose complex problems into subtasks handled by specialized agents. Early frameworks such as CAMEL \citep{li2023camel}, AutoGen \citep{wu2024autogen}, and MetaGPT \citep{hong2023metagpt} demonstrated the potential of collaborative problem-solving through role-playing and structured communication. However, these approaches rely on static, hand-crafted topologies that are labor-intensive to design and difficult to adapt across tasks and difficulty levels.

\paragraph{Automated Multi-Agent Orchestration.} To overcome the rigidity of manual designs, recent work treats orchestration as a search or optimization problem.
AnyMAC \citep{wang2025anymac}, SeqCV \citep{yao2025can}, and Gated-Memory Routing \citep{rajib-etal-2026-learning} dynamically adjust chain length per query.
Aflow \citep{zhang2025aflow} and EvoAgentX \citep{wang2025evoagentx} search over execution topologies using evolutionary or flow-based methods.
MaAS \citep{zhang2025multiagent} models the search space as a probabilistic agentic supernet and trains an orchestrator via policy gradient.
AgentPrune \citep{zhang2025cut} prunes redundant communication messages to obtain more economical communication topologies, while AgentDropout \citep{wang2025agentdropout} dynamically eliminates redundant agents and communication to improve token efficiency.
GDesigner \citep{zhang2024g} uses graph neural networks to architect communication topologies.
These methods primarily optimize task performance, token efficiency, or inference cost. In contrast, latency is a graph-level systems objective whose bottlenecks are concentrated on the critical execution path, making it difficult to optimize with uniform additive rewards alone. INFRAMIND~\citep{kabir2026inframindinfrastructureawaremultiagentorchestration} conditions planning, per-step routing, and scheduling on the runtime state of the serving infrastructure. Orthogonally, systems-level latency management methods such as Kairos~\citep{chen2025kairos}, HW-Router~\citep{hwrouter2026}, and AgentServeSim~\citep{rajib2026agentservesimservingsystemsimulationpolicy} optimize serving, batching, or runtime scheduling, and KV-cache reuse across contexts and model checkpoints~\citep{shi2026kvsharearenakvcachereusecontexts} avoids repeated computation on shared text; both are complementary to our focus on learning latency-aware orchestration policies.

\paragraph{Latent Communication and Agentic Routing.} LatentMAS~\citep{zou2026latentcollaborationmultiagentsystems} is a training-free method that replaces text-based communication among agents with the exchange of latent (hidden-state) representations, reducing decoding and communication overhead. GraphPlanner~\citep{feng2026graphplannergraphmemoryaugmentedagentic} performs agentic routing, selecting an (LLM backbone, agent role) pair at each step to improve accuracy and reduce computational cost, which accumulates additively over the steps of the resulting trajectory. LAMaS differs from both in problem setting, graph construction, execution control, and latency optimization: it samples operator execution graphs and optimizes their critical-path latency under an explicit accuracy floor, assigns latency credit to operators according to path criticality, and adds an inference-time controller that eliminates operators not yet launched. Because an operator off the critical path adds cost but not latency, additive cost and wall-clock latency are distinct objectives; latent communication, in turn, reduces per-operator overhead rather than selecting which operators to execute, and is therefore complementary to our approach. LLM routers such as R2-Router~\citep{pmlr-v306-xue26h}, which jointly select an LLM and an output-length budget for each query, are likewise complementary.

\paragraph{Model-Level Efficiency.} Model compression reduces the cost of each model invocation, for example through mixed-precision quantization~\citep{Lou2020AutoQ:, ijcai2021p472}, weighted low-rank factorization~\citep{hsu2022language}, and dictionary lookups that replace dense weight matrices~\citep{Lou_2022_CVPR, lou2022dictformer}. These model-level techniques are complementary to workflow-level latency optimization such as LAMaS, which targets the critical execution path across agent calls.

\section{Methodology}

\begin{figure}[t]
    \centering
    \includegraphics[width=\textwidth]{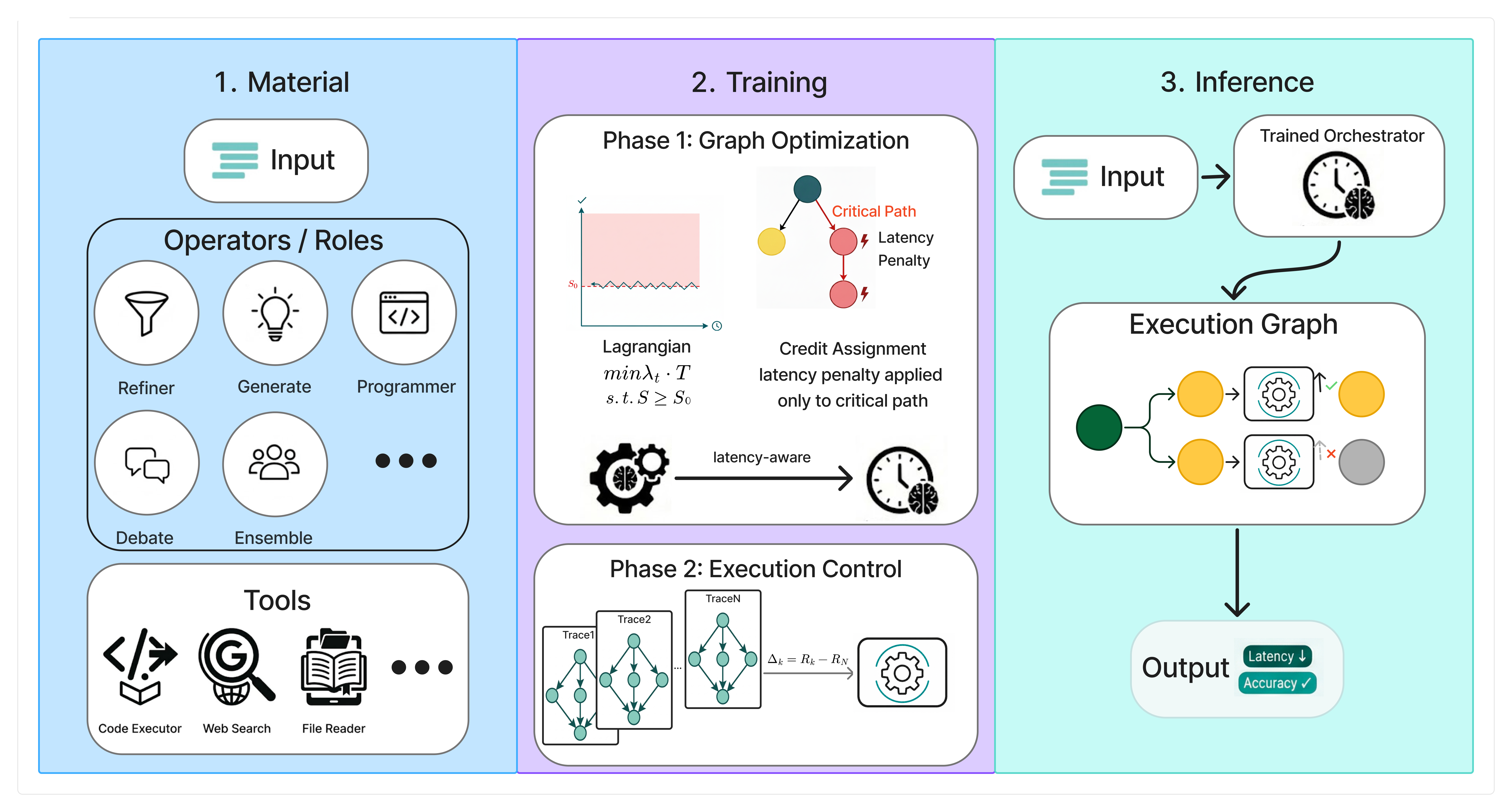}
    \caption{Training and inference in the LAMaS orchestration layer. \textbf{(1)~Material:} A library of reusable operators and tools. \textbf{(2)~Training:} Phase~1 performs latency-aware graph optimization via Lagrangian-constrained learning and critical-path credit assignment, where $S_0$ denotes the target accuracy floor enforced during optimization. Phase~2 trains a lightweight execution controller on the orchestrator's traces to predict the control advantage. \textbf{(3)~Inference:} The trained orchestrator first constructs the execution graph; the execution controller then performs online execution control, monitoring operator completions and adaptively eliminating redundant future agent interactions.}
    \label{fig:framework}
\end{figure}

\subsection{Problem Setting}

We consider a query-dependent multi-agent system (MAS) that dynamically composes a set of \emph{agentic operators} to solve an input problem.
An agentic operator is the basic execution unit, representing a composite agent invocation that may involve multiple LLM calls and external tool usage.
Given a query $x$, the orchestrator constructs an execution graph $\mathcal{G} = (\mathcal{V}, \mathcal{E})$, a directed acyclic graph (DAG) where each node $o \in \mathcal{V}$ is an operator invocation. The edge set $\mathcal{E}$ is induced by operator I/O flow: an edge $(o_i, o_j) \in \mathcal{E}$ exists when the output produced by $o_i$ is used as input by $o_j$ during execution.

We define the \emph{critical path} as the longest path in $\mathcal{G}$ measured by execution time:
\begin{equation}
T = \max_{p \in \text{Paths}(\mathcal{G})} \sum_{o \in p} t(o),
\end{equation}
where $t(o)$ denotes the execution time of operator $o$, and $\text{Paths}(\mathcal{G})$ denotes all source-to-sink paths in $\mathcal{G}$.
The critical path determines the end-to-end latency: operators off the critical path do not affect wall-clock time.
In contrast, the total execution cost accumulates additively across all operators: $C = \sum_{o \in \mathcal{V}} c(o)$.
Latency and cost therefore diverge.

Our goal is to learn an orchestrator that preserves task accuracy while minimizing end-to-end latency.

\paragraph{Execution and final-answer semantics.} Operators in $\mathcal{G}$ produce \emph{candidate solutions} rather than final answers: every completed answer-producing operator contributes its output(s) to a shared candidate pool. The final answer is always emitted by a designated terminal stage (the sink of $\mathcal{G}$) applied to this pool: when several candidates are present, it selects one (e.g., via the self-consistency ensemble); when only one is present, that candidate passes through unchanged. This holds for both full and early-stopped executions. When the execution controller (Section~\ref{sec:gate}) stops an execution early, it only prevents operators that have not been launched from starting; operators already launched finish normally, and the terminal stage then runs once on the resulting candidates.

\subsection{Orchestrator Architecture}
\label{sec:orchestrator}

LAMaS represents the orchestrator as a policy network over an \emph{agentic supernet}---a probabilistic DAG over a fixed set of candidate operators $\mathcal{O}$ (a formulation also adopted by \citet{zhang2025multiagent}). The latency-aware training and credit assignment introduced below are agnostic to the specific architectural choice and apply to any orchestrator that produces a differentiable distribution over execution graphs. The orchestrator constructs $\mathcal{G}$ by sampling operators step-by-step, with the generation probability factorized as:
\begin{equation}
    P_\theta(\mathcal{G}|x) = \prod_{\ell=1}^{L} \pi_\theta(\mathcal{V}_\ell \mid x, \mathcal{G}_{1:\ell-1}),
\end{equation}
where $\mathcal{V}_\ell \subseteq \mathcal{O}$ is the operator subset selected at step $\ell$. The network produces a query-aware operator-level policy via softmax over activation logits; we denote the resulting probability of operator $o$ by $s_o := p_\theta(o \mid x, \mathcal{G}_{1:\ell-1})$. Operators are sampled from this policy sequentially without replacement until their cumulative probability first exceeds a threshold $\tau$:
\begin{equation}
    \mathcal{V}_\ell = \bigl\{o_1, \ldots, o_k\bigr\}, \quad k = \min\!\Bigl\{j : \textstyle\sum_{i=1}^{j} s_{o_i} > \tau\Bigr\},
\end{equation}
enabling the architecture to dynamically adjust width and depth per query.

\subsection{Training-Time Constrained Optimization}

Existing multi-agent orchestration methods typically combine accuracy and cost into a single weighted reward (e.g., $R = S - \lambda_c C$). However, as discussed in Section~1, latency and cost are structurally different objectives. Because latency is governed by the critical path, uniform penalties provide weak operator-level attribution; without an explicit accuracy floor, latency-driven optimization can degrade task performance; and a graph fixed at training time cannot leverage execution-time evidence. LAMaS addresses these via critical-path-aware credit assignment, an accuracy floor enforced by Lagrangian relaxation, and a lightweight execution-time controller. We therefore formulate MAS orchestration as a constrained optimization:
\begin{equation}
\min_\theta \; \mathbb{E}\!\left[\lambda_t T + \lambda_c C\right], \quad \text{s.t.} \quad \mathbb{E}[S] \geq S_0,
\end{equation}
where $S \in \{0,1\}$ is the task score, $C$ is the execution cost in USD per query, $T$ is the end-to-end wall-clock latency in seconds per query, and $\lambda_t$, $\lambda_c$ are fixed weights controlling the relative importance of latency and cost. Here, $S_0$ is a fixed target accuracy floor on $\mathbb{E}[S]$, rather than a per-instance requirement, and in practice is set to the architecture's reference accuracy (see the appendix for sensitivity to different floors).

The accuracy floor $S_0$ acts as a ``do-no-harm'' constraint: LAMaS reduces latency without falling below the reference accuracy of the underlying architecture. In our main experiments, $S_0$ is set to the validation accuracy of the same orchestrator trained with $\lambda_t=0$; we analyze sensitivity to $S_0$ in the appendix. We solve the constrained problem via Lagrangian relaxation with an adaptive dual variable:
\begin{equation}
\label{eq:lagrangian}
R = \lambda_S \frac{S}{S_0} - \lambda_t T - \lambda_c C,
\end{equation}
where $\lambda_S$ is the dual variable updated via projected gradient ascent:
\begin{equation}
\label{eq:dual_update}
\lambda_S \leftarrow \max\!\left(0,\; \lambda_S + \eta \left(1 - \frac{\bar{S}}{S_0}\right)\right),
\end{equation}
where $\bar{S}$ denotes the batch-averaged accuracy and $\eta$ is the dual learning rate.
When accuracy drops below $S_0$, $\lambda_S$ automatically increases to prioritize correctness; when accuracy is satisfied, $\lambda_S$ decreases to allow more aggressive latency reduction. This adaptive coupling removes the need to hand-tune a fixed accuracy--latency trade-off weight.

\subsection{Critical-Path-Aware Credit Assignment}
\label{sec:credit}

End-to-end latency is determined by the critical path. Applying the latency penalty uniformly to all operators creates a mismatch in operator-level credit assignment: operators that have little or no influence on $T$ receive the same gradient as bottleneck operators on the critical path. We therefore use a soft bottleneck-aware weighting that preserves full penalty on the realized critical path while assigning attenuated latency credit to operators on near-critical paths.

For each operator $o$ in the realized execution graph $\mathcal{G}$, let $\ell(o)$ denote the latency of the longest path passing through $o$, and define the path-criticality weight
\begin{equation}
\label{eq:cp_weight}
w(o) \;=\; \frac{\ell(o)}{T} \;\in\; [0, 1].
\end{equation}
By construction $w(o) = 1$ exactly when $o$ lies on the critical path, $w(o) \approx 1$ for operators on near-critical paths, and $w(o) \ll 1$ for operators whose longest containing path is far below $T$. We assign operator-level rewards as
\begin{equation}
\label{eq:credit}
R(o) \;=\; \lambda_S \frac{S}{S_0} \;-\; \lambda_t \, w(o) \, T \;-\; \lambda_c C,
\end{equation}
so each operator receives a fraction $w(o)$ of the global latency penalty $-\lambda_t T$: critical-path operators are penalized in full, near-critical operators receive a smoothly attenuated share, and operators with negligible $w(o)$ get effectively no latency signal. This biases the gradient toward bottleneck operators while preserving softer latency attribution for near-critical paths; the values $\ell(o)$, and hence $w(o)$, are obtained alongside the critical path itself via a single forward/backward pass over $\mathcal{G}$.

\subsection{Learning Objective}

The orchestrator is trained with policy gradient optimization. Let $\xi$ denote the trajectory of operator selections; the loss is
\begin{equation}
\label{eq:policy_loss}
\mathcal{L}(\theta) = - \mathbb{E}_{\xi \sim \pi_\theta} \left[ \sum_{\ell=1}^{L} \sum_{o \in \mathcal{V}_\ell} R(o) \log p_\theta(o \mid x, \mathcal{G}_{1:\ell-1}) \right].
\end{equation}
To reduce variance, we normalize rewards using exponential moving averages (EMA). Specifically, we maintain running estimates $\hat{\mu}$ and $\hat{\sigma}^2$ of the mean and variance of $R(o)$ with momentum 0.95, and normalize each reward as $(R(o) - \hat{\mu}) / \hat{\sigma}$ before computing the policy gradient; this is more stable than per-batch normalization for small batch sizes.

\subsection{Inference-Time Execution Control}
\label{sec:gate}

Because the orchestrator can activate multiple operators per step (Section~\ref{sec:orchestrator}), multiple candidate outputs become available as an execution proceeds, so answer consensus among completed candidates is well-defined at intermediate decision points.
The orchestrator commits to the full execution graph based solely on the query $x$, but each operator completion reveals new information---output content, emerging answer consensus, and elapsed time---unavailable at construction time~\citep{rajib-etal-2026-learning}. LAMaS therefore augments graph construction with an execution-time controller, implemented as a lightweight learned module. This controller can be viewed as performing an online elimination of redundant future agent interactions: at each operator completion, it evaluates whether the current partial execution should be retained or whether additional agent interactions remain worthwhile, by predicting the \emph{control advantage}---how much better (or worse) the current partial execution would be compared to running all operators:
\begin{equation}
\label{eq:control}
\Delta_k = R_k - R_N,
\end{equation}
where $R_k$ is the Lagrangian reward (Eq.~\ref{eq:lagrangian}) evaluated at the $k$-th operator completion and $R_N$ is the reward under full execution.
$R_k$ is a \emph{counterfactual} stop-at-$k$ evaluation: the reward that would be obtained if execution stopped right after the $k$-th operator completion, i.e., if the final answer were produced from the candidates available at that point. It is computed only offline, from the full-execution traces recorded for controller training (described below); at inference time $R_k$ is never computed, and a final answer is produced exactly once, when execution actually stops.
$\Delta_k > 0$ means the current partial execution is preferable because it preserves accuracy while reducing latency; $\Delta_k < 0$ means additional agent interactions remain beneficial.

The controller is a small MLP that takes execution-state features as input and predicts $\hat{\Delta}_k$:
\begin{equation}
\hat{\Delta}_k = \text{Controller}_\phi\!\left(\frac{n_{\text{completed}}}{n_{\text{total}}},\; \frac{\max\_\text{agree}}{n_{\text{cand}}},\; \text{Compress}(q)\right),
\end{equation}
where $n_{\text{completed}}$ is the number of operators that have finished execution at the current decision point, $n_{\text{total}}$ is the total number of operators scheduled in the sampled execution graph, $n_{\text{cand}}$ is the number of candidate answers extracted from the completed operators, and $\max\_\text{agree}$ is the size of the largest cluster of mutually agreeing answers among these candidates. The first feature thus measures execution progress and the second the current rate of agreement (answer consensus); $\text{Compress}(q)$ is a learned low-dimensional projection of the query embedding for difficulty awareness. We define consensus via task-specific semantic equivalence of the current operator outputs: for GSM8K and MATH, we extract the boxed answer (or, if absent, the final numeric answer) and compare exact matches; for MMLU-Pro, we extract the predicted option letter and compare exact matches; for HumanEval, we regard two outputs as agreeing if they induce the same pass/fail outcomes on the test cases available during execution (i.e., the same execution feedback used by the \textbf{Test} operator).
The predicted advantage is converted to a selection score via:
\begin{equation}
p(\text{select current}) = \sigma(\hat{\Delta}_k / \sigma_\Delta),
\end{equation}
where $\sigma_\Delta = \text{std}(\{\Delta_k \mid \Delta_k > 0\})$ is computed from training data. At inference time, we sample the retain/eliminate decision from this Bernoulli score at each decision point.

\paragraph{Controller training.}
The execution controller is trained \emph{after} the orchestrator converges, using execution traces collected by running the frozen orchestrator on the training set.
For each query, all operators are executed to completion, and $\Delta_k$ is computed for every decision point $k$.
The controller is trained via MSE regression: $\min_\phi \mathbb{E}[(\hat{\Delta}_k - \Delta_k)^2]$.
Since $R_k$ uses the same Lagrangian reward as the orchestrator, the controller inherits the orchestrator's latency-aware objective under the same accuracy floor---in particular, the dual variable $\lambda_S$ encodes how difficult accuracy is to maintain, automatically calibrating the controller's aggressiveness.

The full training and inference procedure is summarized in Appendix~\ref{app:full_procedure} (Algorithm~\ref{alg:lamas}). Phase~1 trains the orchestrator with Lagrangian-constrained policy gradient and critical-path credit assignment. Phase~2 trains the execution controller on traces from the frozen orchestrator to predict the control advantage.

\section{Experiments}

\subsection{Experimental Setup}

\paragraph{Benchmarks \& Tasks.} We evaluate our method on four benchmarks spanning three task categories. For \textsc{HumanEval}, \textsc{GSM8K}, and \textsc{MATH}, we follow the same training and evaluation splits as MaAS \citep{zhang2025multiagent}; for \textsc{MMLU-Pro} \citep{wang2024mmlu}, we construct our own stratified train/test split from the official test set. Detailed split statistics are provided in Appendix~\ref{app:split_stats}.
For \textbf{code generation}, we use \textsc{HumanEval}, which measures functional correctness of Python programs generated from natural language descriptions.
For \textbf{mathematical reasoning}, we use \textsc{GSM8K} and \textsc{MATH}, which consist of grade-school and competition-level math problems requiring multi-step reasoning.
For \textbf{knowledge reasoning}, we use \textsc{MMLU-Pro}, which covers diverse professional-level questions requiring multi-step reasoning.

\paragraph{Baselines.} We compare against two categories of baselines: (i) \textbf{non-MAS prompting strategies}---\textbf{Generate} (direct single-step answer), \textbf{CoT}~\citep{wei2022chain}, and \textbf{SC}~\citep{wang2022self}; and (ii) \textbf{learned MAS orchestrators}---\textbf{GDesigner}~\citep{zhang2024g}, \textbf{AgentDropout}~\citep{wang2025agentdropout}, and \textbf{MaAS}~\citep{zhang2025multiagent}. Detailed per-baseline configurations are provided in Appendix~\ref{app:baseline_setup}.

\paragraph{Evaluation Metrics.} We evaluate all methods using three metrics: task performance, API cost, and latency.
For task performance, we report pass@1 on \textsc{HumanEval}, and accuracy on \textsc{GSM8K}, \textsc{MATH}, and \textsc{MMLU-Pro}.
API cost is measured as the average per-query monetary cost (in units of $\times 10^{-4}$ USD) incurred by LLM API calls.
For latency, we directly measure the end-to-end wall-clock execution time, reported as the average latency per query (in seconds).
All methods in Tables~\ref{tab:baseline_comparison}--\ref{tab:generalizability} that use API-served backbones are evaluated under the same API provider and execution runtime, without response caching, and latency is averaged over 3 independent runs.
The end-to-end latency is determined by the critical execution path as defined in Section~3.1.
We also report CP Tokens, the average number of output tokens generated along the critical execution path, as a complementary structural indicator of critical-path workload.

\paragraph{Implementation Details.}
We use the closed-source LLM \texttt{gpt-4o-mini-0718} \citep{openai2024gpt4omini}. The latency penalty is set to $\lambda_t=0.005$, the activation threshold to $\tau=0.3$, and the supernet has $L=4$ layers; we report sensitivity to each of these in the appendix. The target accuracy floor $S_0$ is set to the validation accuracy of the same orchestrator trained without latency optimization (i.e., $\lambda_t=0$), so the constraint measures whether latency reduction is achieved while maintaining the reference accuracy level of the underlying architecture. Remaining training hyperparameters are listed in Appendix~\ref{app:impl_details}. All results are averaged over 3 independent runs.

\begin{table}[t]
\centering
\small
\caption{Comparison with non-MAS and MAS baselines across four benchmarks. Each cell reports mean $\pm$ standard deviation across 3 independent runs. For MAS methods, the best mean in each column per dataset is shown in \textbf{bold}. Cost is reported in units of $\times 10^{-4}$ USD. Latency is the average end-to-end wall-clock time per query (in seconds). CP Tokens denotes the average number of output tokens on the critical execution path.}
\label{tab:baseline_comparison}
\begin{tabular}{llcccc}
\toprule
Dataset & Method & Acc (\%) $\uparrow$ & Cost ($\times 10^{-4}$) $\downarrow$ & Latency (s) $\downarrow$ & CP Tokens $\downarrow$ \\
\midrule
\multirow{7}{*}{GSM8K}
& Generate & $87.46_{\pm 0.20}$ & $2.87_{\pm 0.05}$ & $5.43_{\pm 0.18}$ & $280_{\pm 8}$ \\
& CoT & $88.97_{\pm 0.43}$ & $4.53_{\pm 0.04}$ & $6.92_{\pm 0.21}$ & $331_{\pm 15}$ \\
& SC & $89.86_{\pm 0.19}$ & $18.61_{\pm 0.09}$ & $10.64_{\pm 0.27}$ & $591_{\pm 22}$ \\
\cmidrule{2-6}
& GDesigner & $93.30_{\pm 0.93}$ & $17.86_{\pm 0.28}$ & $32.31_{\pm 1.07}$ & $1624_{\pm 29}$ \\
& AgentDropout & $93.11_{\pm 0.27}$ & $53.81_{\pm 6.78}$ & $81.24_{\pm 12.74}$ & $4098_{\pm 538}$ \\
& MaAS & $93.36_{\pm 0.72}$ & $31.52_{\pm 0.78}$ & $48.11_{\pm 3.49}$ & $2501_{\pm 53}$ \\
& \textbf{LAMaS} & $\mathbf{93.65_{\pm 0.41}}$ & $\mathbf{7.71_{\pm 0.94}}$ & $\mathbf{11.73_{\pm 1.13}}$ & $\mathbf{622_{\pm 87}}$ \\
\midrule
\multirow{7}{*}{HumanEval}
& Generate & $84.73_{\pm 1.32}$ & $5.11_{\pm 0.06}$ & $10.23_{\pm 0.71}$ & $513_{\pm 31}$ \\
& CoT & $85.50_{\pm 3.05}$ & $7.24_{\pm 0.32}$ & $14.87_{\pm 1.04}$ & $780_{\pm 47}$ \\
& SC & $86.26_{\pm 0.76}$ & $28.15_{\pm 0.13}$ & $20.41_{\pm 0.57}$ & $1105_{\pm 38}$ \\
\cmidrule{2-6}
& GDesigner & $85.50_{\pm 2.02}$ & $19.86_{\pm 2.46}$ & $49.68_{\pm 5.62}$ & $2983_{\pm 489}$ \\
& AgentDropout & $86.01_{\pm 2.20}$ & $27.73_{\pm 1.32}$ & $51.49_{\pm 6.93}$ & $3043_{\pm 218}$ \\
& MaAS & $93.38_{\pm 1.17}$ & $28.18_{\pm 1.21}$ & $47.32_{\pm 2.93}$ & $2708_{\pm 109}$ \\
& \textbf{LAMaS} & $\mathbf{95.42_{\pm 1.53}}$ & $\mathbf{12.92_{\pm 1.16}}$ & $\mathbf{21.00_{\pm 2.84}}$ & $\mathbf{1180_{\pm 187}}$ \\
\midrule
\multirow{7}{*}{MATH}
& Generate & $50.62_{\pm 1.09}$ & $6.38_{\pm 0.14}$ & $10.17_{\pm 0.34}$ & $491_{\pm 19}$ \\
& CoT & $51.23_{\pm 1.15}$ & $10.24_{\pm 0.07}$ & $21.03_{\pm 0.31}$ & $992_{\pm 21}$ \\
& SC & $51.85_{\pm 1.23}$ & $42.41_{\pm 0.27}$ & $27.62_{\pm 0.41}$ & $1340_{\pm 24}$ \\
\cmidrule{2-6}
& GDesigner & $50.82_{\pm 0.71}$ & $30.17_{\pm 2.27}$ & $60.71_{\pm 5.83}$ & $3155_{\pm 246}$ \\
& AgentDropout & $52.47_{\pm 0.94}$ & $97.23_{\pm 8.61}$ & $176.42_{\pm 19.18}$ & $8693_{\pm 974}$ \\
& MaAS & $52.67_{\pm 0.82}$ & $65.39_{\pm 1.19}$ & $109.87_{\pm 5.92}$ & $5602_{\pm 116}$ \\
& \textbf{LAMaS} & $\mathbf{53.29_{\pm 1.25}}$ & $\mathbf{28.74_{\pm 3.05}}$ & $\mathbf{28.41_{\pm 4.27}}$ & $\mathbf{1452_{\pm 232}}$ \\
\midrule
\multirow{7}{*}{MMLU-Pro}
& Generate & $54.47_{\pm 1.14}$ & $1.05_{\pm 0.08}$ & $2.68_{\pm 0.22}$ & $136_{\pm 7}$ \\
& CoT & $56.80_{\pm 3.03}$ & $2.41_{\pm 0.15}$ & $6.27_{\pm 0.46}$ & $361_{\pm 14}$ \\
& SC & $58.67_{\pm 1.21}$ & $14.88_{\pm 0.18}$ & $10.53_{\pm 0.49}$ & $620_{\pm 12}$ \\
\cmidrule{2-6}
& GDesigner & $64.27_{\pm 3.63}$ & $25.03_{\pm 0.42}$ & $46.66_{\pm 2.47}$ & $2750_{\pm 41}$ \\
& AgentDropout & $61.87_{\pm 2.31}$ & $76.13_{\pm 1.41}$ & $119.33_{\pm 14.68}$ & $6560_{\pm 561}$ \\
& MaAS & $65.40_{\pm 3.08}$ & $30.37_{\pm 2.56}$ & $49.12_{\pm 5.13}$ & $2985_{\pm 151}$ \\
& \textbf{LAMaS} & $\mathbf{66.27_{\pm 1.96}}$ & $\mathbf{10.93_{\pm 1.39}}$ & $\mathbf{14.81_{\pm 1.92}}$ & $\mathbf{845_{\pm 96}}$ \\
\bottomrule
\end{tabular}
\end{table}

\subsection{Result Analysis}

We analyze the results from three perspectives: evaluated learning-based MAS baselines, the full set of baselines, and component-wise ablations.

\paragraph{Comparison with evaluated learning-based MAS baselines.}
Table~\ref{tab:baseline_comparison} shows that LAMaS consistently reduces end-to-end latency by 55.6--75.6\% relative to the strongest evaluated learning-based MAS baselines while maintaining competitive or better accuracy. The reduction co-occurs with substantially lower CP Tokens (e.g., $622$ vs.\ $2501$ on GSM8K), indicating that the orchestrator produces graphs with shorter critical paths under the latency objective; the inference-time controller further trims execution once a confident answer emerges.

\paragraph{Comparison with baselines.}
Non-MAS methods (\textbf{Generate}, \textbf{CoT}, \textbf{SC}) remain faster because of their simplicity but consistently underperform in task accuracy. Among MAS methods, \textbf{GDesigner} and \textbf{AgentDropout} incur substantially higher latency without corresponding accuracy gains, whereas LAMaS achieves the lowest latency across all four benchmarks while maintaining comparable accuracy. Across MAS methods and benchmarks, CP Tokens are strongly correlated with measured wall-clock latency (Appendix~\ref{app:cp_token_correlation}), providing complementary evidence that LAMaS reduces critical-path workload together with end-to-end latency. Appendix~\ref{app:single_agent} further compares LAMaS with single-agent agentic baselines that add tool use or self-correction (tool-augmented CoT, test-time repair, and Self-Refine); LAMaS remains 3.86--9.14~pp more accurate than these baselines.

\subsection{Ablation Study}

We study four variants of LAMaS: removing CP credit assignment, removing the inference-time controller, removing the latency objective, and a naive ``add latency to the reward'' baseline. \emph{w/o CP Credit} forces the latency penalty to apply uniformly to all operators rather than weighting it by path criticality; \emph{w/o controller} removes the execution controller entirely and relies solely on training-time optimization; \emph{w/o latency obj.} sets $\lambda_t=0$, training the orchestrator without any latency objective; \emph{naive lat. only} keeps only a bare uniform latency term in the training objective, with no $S_0$ accuracy floor, no CP-aware credit assignment, and no inference controller---the simplest ``add latency to the reward'' baseline.

\begin{table}[t]
\centering
\small
\caption{Ablation study of inference-time execution control, critical-path (CP) credit assignment, and the latency objective across four benchmarks. Cost is in units of $\times 10^{-4}$ USD; latency is end-to-end wall-clock time per query (in seconds).}
\label{tab:ablation}
\begin{tabular}{llccc}
\toprule
Dataset & Method & Acc (\%) $\uparrow$ & Cost ($\times 10^{-4}$) $\downarrow$ & Latency (s) $\downarrow$ \\
\midrule
\multirow{5}{*}{GSM8K}
& LAMaS & \textbf{93.65} & \textbf{7.71} & \textbf{11.73} \\
& w/o controller & 93.27 & 9.43 & 13.84 \\
& w/o CP Credit & 92.73 & 8.71 & 12.93 \\
& w/o latency obj. ($\lambda_t{=}0$) & 93.59 & 9.71 & 15.83 \\
& naive lat. only & 91.85 & 9.86 & 15.52 \\
\midrule
\multirow{5}{*}{HumanEval}
& LAMaS & 95.42 & \textbf{12.92} & \textbf{21.00} \\
& w/o controller & 95.17 & 16.18 & 24.47 \\
& w/o CP Credit & 94.40 & 14.83 & 23.43 \\
& w/o latency obj. ($\lambda_t{=}0$) & \textbf{95.67} & 17.42 & 35.62 \\
& naive lat. only & 93.13 & 18.07 & 31.84 \\
\midrule
\multirow{5}{*}{MATH}
& LAMaS & \textbf{53.29} & \textbf{28.74} & \textbf{28.41} \\
& w/o controller & 53.02 & 34.51 & 33.46 \\
& w/o CP Credit & 52.26 & 32.18 & 31.27 \\
& w/o latency obj. ($\lambda_t{=}0$) & 53.16 & 36.84 & 35.27 \\
& naive lat. only & 51.99 & 37.21 & 34.27 \\
\midrule
\multirow{5}{*}{MMLU-Pro}
& LAMaS & 66.27 & \textbf{10.93} & \textbf{14.81} \\
& w/o controller & 66.00 & 12.68 & 17.94 \\
& w/o CP Credit & 65.40 & 11.83 & 16.18 \\
& w/o latency obj. ($\lambda_t{=}0$) & \textbf{66.33} & 13.05 & 22.46 \\
& naive lat. only & 60.93 & 13.74 & 20.91 \\
\bottomrule
\end{tabular}
\end{table}

Table~\ref{tab:ablation} shows consistent trends across all four benchmarks. Removing the controller preserves accuracy within 0.38~pp but increases latency by 16.5--21.1\% and cost by 16.0--25.2\%, confirming the importance of inference-time execution control. Removing CP credit assignment also degrades all three metrics, supporting path-criticality-weighted latency attribution rather than a uniform penalty over all operators. Removing the latency objective entirely ($\lambda_t{=}0$) leaves accuracy essentially unchanged (within $\pm 0.25$~pp of LAMaS) but inflates latency by 24--70\% and cost by 19--35\%, confirming that the constrained latency objective reshapes the execution graph for efficiency while maintaining the reference accuracy level of the underlying architecture. Finally, the \emph{naive lat. only} variant drops accuracy by 1.30--5.34~pp relative to LAMaS while yielding only 2--11\% latency savings over $\lambda_t{=}0$ and no cost reduction, showing that simply adding a latency term provides only limited gains. Taken together, these ablations show that LAMaS combines path-selective latency pressure with an explicit accuracy floor and execution-time control rather than simply shrinking computation.

\subsection{Transferability and Generality of LAMaS Mechanisms}

The full LAMaS pipeline---the training-time graph optimization, the critical-path credit assignment, and the inference-time execution controller---is not tied to a specific orchestrator architecture. To verify its broader applicability, we integrate the entire pipeline into two other learning-based MAS methods, GDesigner \citep{zhang2024g} and AgentDropout \citep{wang2025agentdropout}, without redesigning their original orchestration frameworks.

\begin{table}[!t]
\centering
\small
\caption{Transferability of the full LAMaS pipeline (training-time graph optimization, critical-path credit assignment, and inference-time execution control) to other learning-based MAS architectures. The controller is retrained on each architecture's own traces (Phase~2 of the LAMaS training procedure). Latency is the average end-to-end wall-clock time per query (in seconds). CP Tokens denotes the average critical-path output tokens.}
\label{tab:generalizability}
\begin{tabular}{llcccc}
\toprule
Dataset & Method & Acc (\%) $\uparrow$ & Cost ($\times 10^{-4}$) $\downarrow$ & Latency (s) $\downarrow$ & CP Tokens $\downarrow$ \\
\midrule
\multirow{4}{*}{GSM8K}
& GDesigner & 93.30 & 17.86 & 32.31 & 1624 \\
& \textbf{GDesigner+LAMaS} & \textbf{93.27} & \textbf{12.08} & \textbf{20.55} & \textbf{1138} \\
\cmidrule{2-6}
& AgentDropout & 93.11 & 53.81 & 81.24 & 4098 \\
& \textbf{AgentDropout+LAMaS} & \textbf{93.18} & \textbf{39.67} & \textbf{58.14} & \textbf{2989} \\
\midrule
\multirow{4}{*}{HumanEval}
& GDesigner & 85.50 & 19.86 & 49.68 & 2983 \\
& \textbf{GDesigner+LAMaS} & \textbf{85.75} & \textbf{12.93} & \textbf{30.85} & \textbf{1571} \\
\cmidrule{2-6}
& AgentDropout & 86.01 & 27.73 & 51.49 & 3043 \\
& \textbf{AgentDropout+LAMaS} & \textbf{86.26} & \textbf{17.98} & \textbf{36.62} & \textbf{2200} \\
\midrule
\multirow{4}{*}{MATH}
& GDesigner & 50.82 & 30.17 & 60.71 & 3155 \\
& \textbf{GDesigner+LAMaS} & \textbf{51.23} & \textbf{20.03} & \textbf{39.93} & \textbf{2118} \\
\cmidrule{2-6}
& AgentDropout & 52.47 & 97.23 & 176.42 & 8693 \\
& \textbf{AgentDropout+LAMaS} & \textbf{52.40} & \textbf{70.35} & \textbf{128.34} & \textbf{6311} \\
\midrule
\multirow{4}{*}{MMLU-Pro}
& GDesigner & 64.27 & 25.03 & 46.66 & 2750 \\
& \textbf{GDesigner+LAMaS} & \textbf{64.33} & \textbf{17.03} & \textbf{28.27} & \textbf{1548} \\
\cmidrule{2-6}
& AgentDropout & 61.87 & 76.13 & 119.33 & 6560 \\
& \textbf{AgentDropout+LAMaS} & \textbf{62.20} & \textbf{56.23} & \textbf{82.27} & \textbf{4628} \\
\bottomrule
\end{tabular}
\end{table}

Across all four benchmarks and both architectures, integrating the full LAMaS pipeline consistently reduces latency by 27--39\% while preserving or slightly improving accuracy within typical 3-run noise. These results confirm that latency-aware graph optimization together with inference-time execution control transfers effectively to other learning-based MAS architectures. The reductions are smaller than in the LAMaS-native setting (Table~\ref{tab:baseline_comparison}), consistent with the fact that GDesigner and AgentDropout retain their original orchestration policies: the LAMaS pipeline contributes the latency-aware credit signal and inference-time control rather than replacing their graph construction. Together with Appendix~\ref{app:qwen}, this suggests that the LAMaS pipeline remains effective beyond a single orchestration architecture and extends to the additional backbone and serving configurations we evaluate. Appendix~\ref{app:unseen_operator} further shows that the orchestrator can still reasonably select an operator not exposed during training. Appendix~\ref{app:naive_rules} additionally compares GDesigner+LAMaS with two naive inference-time latency-cutting rules applied to GDesigner; neither rule matches its accuracy--latency trade-off.
\vspace{-2mm}

\FloatBarrier

\section{The LAMaS Platform}
\label{sec:platform}

An application can use LAMaS through its OpenAI-compatible endpoint by changing the server address (Figure~\ref{fig:platform}). For each query, the orchestrator selects a graph of operators. The runtime starts each operator when its inputs are ready and returns the final answer with a trace of operator timings, the critical path, and cost. Operators and graph builders are replaceable; model calls can use a local server such as vLLM or a hosted OpenAI-compatible API.

A framework that builds its own graphs can reuse the orchestration layer directly. In GDesigner and AgentDropout, we train the existing graph policy with the accuracy constraint and critical-path credit. Once that policy is frozen, its execution traces supply the examples for training the stop controller. This preserves each framework's graph construction while letting partial results determine whether the remaining operators should run (Table~\ref{tab:generalizability}).

\section{Conclusion}

LAMaS makes low-latency orchestration reusable across multi-agent systems. Its training objective targets the critical path under an accuracy floor, and its controller uses partial results to skip future calls. Across four benchmarks, LAMaS reduces latency by over 50\% relative to the evaluated learning-based MAS baselines with competitive or better accuracy. GDesigner and AgentDropout also reduce latency while retaining their own graph builders.

\begin{ack}
This work was supported in part by the National Science
Foundation under Award No. 2544047. 
Any opinions, findings, and conclusions or recommendations expressed in this material are those of the author(s) and do not
necessarily reflect the views of the National Science Foundation.
\end{ack}

\clearpage
\bibliographystyle{plainnat}
\bibliography{custom}

\appendix

\section{Limitations}

LAMaS reduces orchestration latency; however, it does not mitigate broader risks inherited from the underlying LLMs and associated tools, including unsafe outputs, hallucinations, and social biases. 

\section{Experimental Setup Details}

This section collects additional implementation and dataset details used for reproducibility.

\subsection{Operator Set}
\label{app:operator_set}

Each operator corresponds to a predefined reasoning or execution primitive used to construct execution graphs.
We briefly summarize their functionality below for completeness.

\begin{itemize}
    \item \textbf{Generate.}
    A basic generator that directly produces text or code without additional reasoning or post-processing.
    It invokes the LLM once and is primarily used for simple generation tasks.

    \item \textbf{GenerateCoT.}
    A chain-of-thought generator that prompts the LLM to perform step-by-step reasoning.
    For mathematical tasks (e.g., MATH, GSM8K), it includes explicit reasoning exemplars, while for code tasks (e.g., HumanEval), it uses lightweight reasoning prompts.
    This operator invokes the LLM once.

    \item \textbf{MultiGenerateCoT.}
    A diversity-oriented CoT generator that produces multiple candidate solutions in parallel.
    It generates three independent chain-of-thought solutions, yielding a set of candidate responses for downstream aggregation.

    \item \textbf{ScEnsemble.}
    A self-consistency ensemble operator that selects the most consistent answer from multiple candidate solutions.
    All candidates are formatted as discrete options, and the LLM is prompted to select the most consistent one, following the self-consistency principle.

    \item \textbf{SelfRefine.}
    A refinement operator that analyzes an existing solution to identify errors or suboptimal reasoning and generates an improved version.
    This operator invokes the LLM once.

\end{itemize}

\paragraph{Task-Specific Operators.}
We also employ task-specific operators for different benchmarks.

\begin{itemize}
    \item \textbf{CustomCodeGenerate} (HumanEval).
    A lightweight code generator that produces candidate code solutions without execution or testing.

    \item \textbf{Test} (HumanEval).
    A test-driven refinement operator that executes generated code and iteratively improves it based on failure feedback.
    Upon failure, the operator generates a revised solution using reflective prompts and retries up to three times.

    \item \textbf{Programmer} (MATH/GSM8K).
    A code execution operator that generates Python programs, executes them in an isolated environment, and iteratively refines the code based on execution feedback.
\end{itemize}

\subsection{Additional Implementation Details}
\label{app:impl_details}

We list the remaining training hyperparameters not detailed in Section~4.1. The cost weight is $\lambda_c=3.0$, the dual learning rate is $\eta=0.01$, the batch size is $4$, and the orchestrator is trained for $4$ repetitions over the training set. These values follow the supernet implementation conventions of MaAS~\citep{zhang2025multiagent}.

Queries are encoded by a frozen \texttt{all-MiniLM-L6-v2} SentenceTransformer; the resulting embedding is shared between the orchestrator's policy network (Section~\ref{sec:orchestrator}) and the controller's $\text{Compress}(\cdot)$ projection (Section~\ref{sec:gate}), and is not finetuned during training.

In the main experiments, we use \texttt{gpt-4o-mini-0718} via the OpenAI API with temperature set to 1. Appendix~\ref{app:qwen} instead evaluates \texttt{qwen3.5-flash} (closed-source) via OpenRouter with \texttt{Alibaba Cloud Int.} selected as the provider, and \texttt{qwen3.5-9b} (open-weight) served locally through vLLM; the same temperature setting is used there as well.

Training the orchestrator and execution controller requires modest local compute but does not finetune the underlying LLM backbone. All training runs were executed on a Slurm cluster using a single NVIDIA H100 80GB GPU per run. The local \texttt{qwen3.5-9b} evaluation uses vLLM served on two NVIDIA B200 GPUs.

\subsection{Dataset Split Statistics}
\label{app:split_stats}

We evaluate on four publicly available benchmarks: \textsc{HumanEval}~\citep{chen2021evaluating}, \textsc{GSM8K}~\citep{cobbe2021training}, \textsc{MATH}~\citep{hendrycks2021measuring}, and \textsc{MMLU-Pro}~\citep{wang2024mmlu}. Following common practice in workflow-automation work \citep{zhang2025multiagent, zhang2025aflow}, each benchmark is partitioned into training and test sets with a train:test ratio of 1:4. For \textsc{MATH}, we use a 605-problem subset of level-5 problems spanning four representative categories: Combinatorics \& Probability, Number Theory, Pre-algebra, and Pre-calculus. For \textsc{MMLU-Pro}, we construct a stratified split from the official test set (seed 42), sampling proportionally per category with no overlap between train and test. Statistics are summarized in Table~\ref{tab:split_stats}.

\begin{table}[h]
\centering
\small
\caption{Dataset statistics.}
\label{tab:split_stats}
\begin{tabular}{llccc}
\toprule
Domain & Dataset & \#Train & \#Test & Metric \\
\midrule
Code Generation & HumanEval & 33 & 131 & pass@1 \\
\midrule
\multirow{2}{*}{Math Reasoning} & GSM8K & 264 & 1055 & Accuracy \\
 & MATH & 119 & 486 & Accuracy \\
\midrule
Knowledge Reasoning & MMLU-Pro & 125 & 500 & Accuracy \\
\bottomrule
\end{tabular}
\end{table}

\subsection{Baseline Setups}
\label{app:baseline_setup}

We provide the configurations for each baseline method below.

\begin{enumerate}
    \item \textbf{Generate.} A single-pass direct-answer baseline that prompts the LLM once with the query and uses its output as the final answer, with no chain-of-thought scaffolding or post-processing.
    \item \textbf{CoT.} Chain-of-Thought prompting~\citep{wei2022chain} elicits step-by-step reasoning before the final answer; we use a standard zero-shot CoT prompt template.
    \item \textbf{SC.} Self-Consistency~\citep{wang2022self}: to enhance robustness, we aggregate five CoT-generated solutions through an ensemble step over the candidate answers.
    \item \textbf{GDesigner.} We use the official implementation of~\citet{zhang2024g}, which learns a graph neural network over operator embeddings to construct communication topologies. We retain the original training procedure and operator set defined in the released codebase, replacing only the LLM backbone with \texttt{gpt-4o-mini-0718} for fair comparison.
    \item \textbf{AgentDropout.} We follow the released implementation of~\citet{wang2025agentdropout}, which dynamically prunes redundant agents and communication edges during execution. The dropout schedule and per-agent role assignments use the defaults from the official codebase.
    \item \textbf{MaAS.} We use the agentic-supernet implementation of~\citet{zhang2025multiagent}, training the orchestrator with the same supernet depth ($L=4$), operator set, and policy-gradient procedure as released. The original cost-only reward is used unchanged so that MaAS represents the cost-aware---but not latency-aware---learning baseline.
\end{enumerate}

All MAS baselines are trained and evaluated under the same train/test partition (Appendix~\ref{app:split_stats}), the same backbone LLM (\texttt{gpt-4o-mini-0718}), and the same per-query latency/cost measurement protocol described in Section~4.1.

\section{Additional Supporting Analyses}

This section provides supplementary analyses supporting the main-paper interpretation of LAMaS.

\subsection{Critical-Path Tokens as a Latency Proxy}
\label{app:cp_token_correlation}

The critical-path-aware credit assignment in Section~3.4 is motivated by the expectation that CP Tokens should co-vary with end-to-end wall-clock latency. To verify this empirically, we form one $(\text{CP Tokens}, \text{latency})$ pair for each MAS-method/benchmark cell in Table~\ref{tab:baseline_comparison} and compute the pooled correlation across all 4 MAS methods $\times$ 4 benchmarks ($n=16$). Both Pearson and Spearman correlations indicate a strong positive association, providing complementary evidence that reductions in CP Tokens align with reductions in end-to-end latency.

\begin{table}[h]
\centering
\small
\caption{Correlation between CP Tokens and measured wall-clock latency across MAS-method/benchmark cells in Table~\ref{tab:baseline_comparison}.}
\label{tab:cp_token_correlation}
\begin{tabular}{lcc}
\toprule
Setting & Pearson $r$ & Spearman $\rho$ \\
\midrule
MAS methods $\times$ 4 benchmarks & 0.995 & 0.985 \\
\bottomrule
\end{tabular}
\end{table}

\subsection{Comparison with Oracle Controller}
\label{app:oracle}

To probe how close the learned execution controller is to an optimal stopping rule, we compare it against an oracle that has access to the true control advantage $\Delta_k = R_k - R_N$ at each decision point and stops iff $\Delta_k > 0$. We evaluate on HumanEval and MMLU-Pro from two complementary angles: decision-level agreement with the oracle (Table~\ref{tab:oracle_decision}), and end-to-end performance when the learned controller is replaced by either the oracle or no controller (Table~\ref{tab:oracle_performance}).

\paragraph{Decision-level agreement.}
Table~\ref{tab:oracle_decision} reports decision accuracy together with precision, recall, and F1 for the positive (stop) class, treating the oracle's $\Delta_k > 0$ rule as ground truth. Across both benchmarks, the learned controller agrees with oracle decisions in 70.8--73.5\% of cases. Precision exceeds recall on both benchmarks, indicating that the learned controller is conservative: it tends to continue execution when the oracle's signal is borderline rather than making an over-aggressive elimination decision.

\begin{table}[h]
\centering
\small
\caption{Decision-level agreement between the learned execution controller and the oracle stopping rule on HumanEval and MMLU-Pro. The positive class is the ``stop'' decision.}
\label{tab:oracle_decision}
\begin{tabular}{lcccc}
\toprule
Benchmark & Decision Acc (\%) $\uparrow$ & Precision $\uparrow$ & Recall $\uparrow$ & F1 $\uparrow$ \\
\midrule
HumanEval & 73.5 & 0.72 & 0.66 & 0.69 \\
MMLU-Pro  & 70.8 & 0.69 & 0.63 & 0.66 \\
\bottomrule
\end{tabular}
\end{table}

\paragraph{End-to-end performance.}
Table~\ref{tab:oracle_performance} compares three controller settings: no controller, the learned controller, and the oracle. The learned controller closes most of the gap between no-controller execution and the oracle: on HumanEval it trails the oracle by 2.16~s in latency and 0.25~pp in accuracy, and on MMLU-Pro by 1.74~s and 0.46~pp. This indicates that the learned controller approaches oracle-level execution control despite operating only on the partial-execution features available at runtime.

\begin{table}[h]
\centering
\small
\caption{End-to-end accuracy and latency under different controllers on HumanEval and MMLU-Pro. The ``Learned'' rows correspond to the LAMaS configuration in Table~\ref{tab:ablation}, and ``No controller'' rows correspond to the \emph{w/o controller} ablation.}
\label{tab:oracle_performance}
\begin{tabular}{llcc}
\toprule
Benchmark & Controller & Acc (\%) $\uparrow$ & Latency (s) $\downarrow$ \\
\midrule
\multirow{3}{*}{HumanEval}
& No controller    & 95.17 & 24.47 \\
& Learned          & 95.42 & 21.00 \\
& \textbf{Oracle}  & \textbf{95.67} & \textbf{18.84} \\
\midrule
\multirow{3}{*}{MMLU-Pro}
& No controller    & 66.00 & 17.94 \\
& Learned          & 66.27 & 14.81 \\
& \textbf{Oracle}  & \textbf{66.73} & \textbf{13.07} \\
\bottomrule
\end{tabular}
\end{table}

\subsection{Full Training and Inference Procedure}
\label{app:full_procedure}

\begin{algorithm}[h]
\small
\caption{LAMaS: Latency-Aware MAS Orchestration}
\label{alg:lamas}
\begin{algorithmic}[1]
\REQUIRE Orchestrator $\pi_\theta$, operator set $\mathcal{O}$, training queries $\mathcal{D}$, accuracy floor $S_0$
\REQUIRE Hyperparameters: $\lambda_t$, $\lambda_c$, dual learning rate $\eta$, threshold $\tau$
\STATE \textbf{// Phase 1: Train orchestrator}
\STATE Initialize dual variable $\lambda_S \leftarrow 1.0$
\FOR{each training iteration}
    \STATE Sample a batch of queries $\{x_i\}$ from $\mathcal{D}$
    \FOR{each query $x_i$}
        \STATE Construct execution graph $\mathcal{G}_i$ via $\pi_\theta$
        \STATE Execute $\mathcal{G}_i$; observe $S_i, C_i, T_i$, per-operator times $\{t(o)\}$
        \STATE Compute $\ell(o)$ for all operators; compute $R(o)$ via Eq.~\ref{eq:credit}
    \ENDFOR
    \STATE Update $\theta$ via policy gradient (Eq.~\ref{eq:policy_loss})
    \STATE $\lambda_S \leftarrow \max(0,\; \lambda_S + \eta(1 - \bar{S}/S_0))$
\ENDFOR
\STATE \textbf{// Phase 2: Train execution controller}
\STATE Freeze $\pi_\theta$; collect full execution traces on $\mathcal{D}$
\FOR{each trace}
    \FOR{each decision point $k = 1, \ldots, N$}
        \STATE Compute $\Delta_k = R_k - R_N$ (Eq.~\ref{eq:control})
        \STATE Record training sample: $(f_k, \Delta_k)$
    \ENDFOR
\ENDFOR
\STATE Train $\text{Controller}_\phi$ via MSE: $\min_\phi \sum (\hat{\Delta}_k - \Delta_k)^2$
\STATE Compute $\sigma_\Delta = \text{std}(\{\Delta_k \mid \Delta_k > 0\})$
\STATE \textbf{// Inference}
\STATE Construct $\mathcal{G}$ via $\pi_\theta$; launch operators
\FOR{each operator completion $k$}
    \STATE Compute $p(\text{select current})$ from $\text{Controller}_\phi(f_k)$, sample a Bernoulli decision; on a positive draw, eliminate the following unlaunched operators.
\ENDFOR
\RETURN Final answer
\end{algorithmic}
\end{algorithm}

\subsection{Compatibility with Unseen Operators}
\label{app:unseen_operator}

\begin{figure}[h]
\centering
\includegraphics[width=0.72\linewidth]{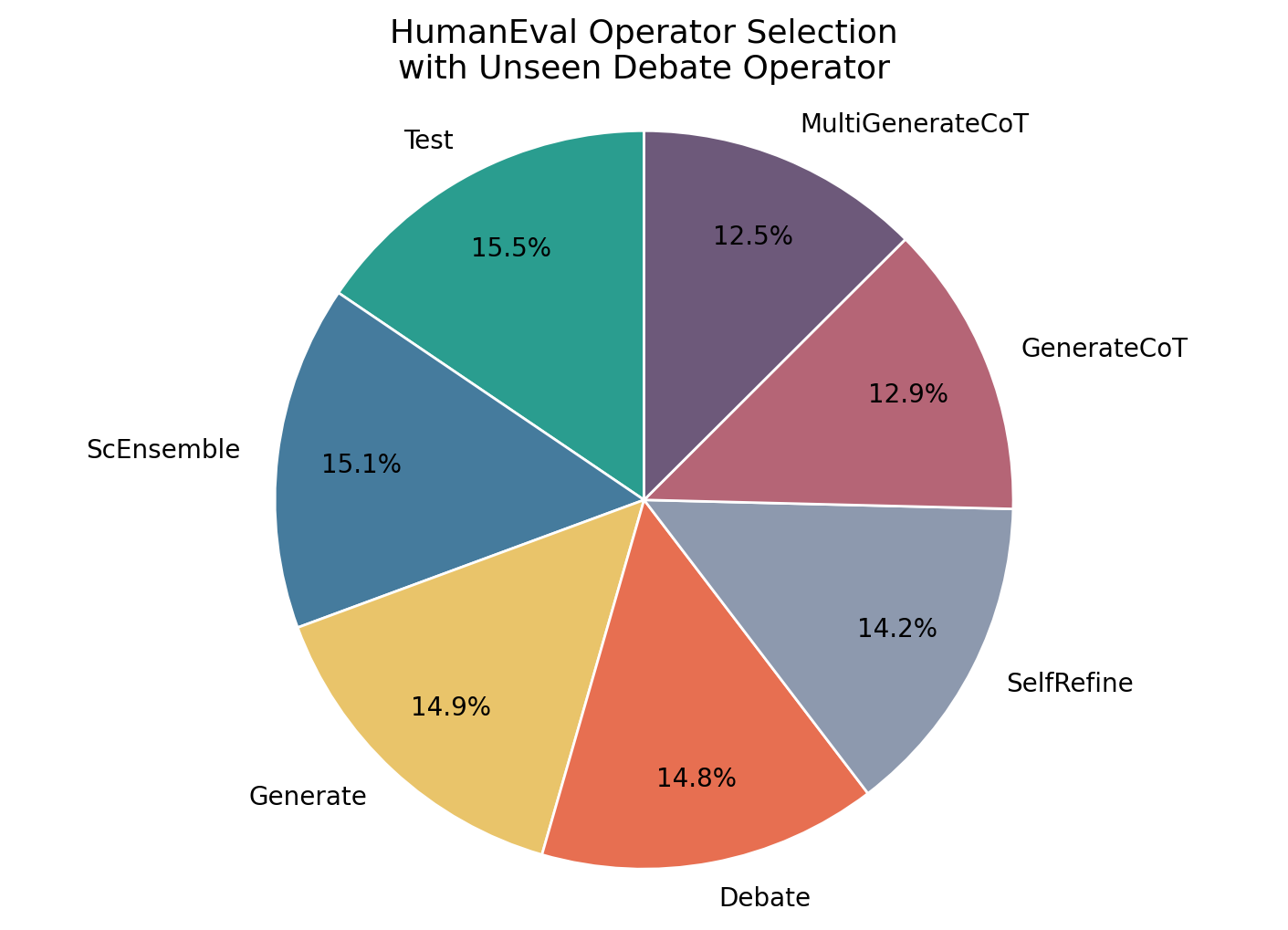}
\caption{Average operator selection probabilities on HumanEval after inserting an unseen operator, \textbf{Debate}, at inference time. The unseen operator receives non-trivial probability mass comparable to several seen operators, indicating that the learned policy can still reasonably select operators not exposed during training.}
\label{fig:unseen_operator_pie}
\end{figure}

To probe whether the learned policy generalizes beyond the fixed training-time operator set, we insert an unseen operator (\textbf{Debate}) at inference time on HumanEval and inspect the orchestrator's average selection probabilities. Figure~\ref{fig:unseen_operator_pie} reports the average probability mass over the seven non-halt operators shown. The unseen \textbf{Debate} operator receives 14.8\% probability mass, comparable to seen operators such as \textbf{Generate} (14.9\%), \textbf{ScEnsemble} (15.1\%), and \textbf{Test} (15.5\%). This suggests that the learned policy can still reasonably select an unseen operator during architecture sampling, indicating some inductive flexibility beyond the exact operator set seen during training.

\section{Sensitivity Analyses}
\label{app:sensitivity}

This section reports the sensitivity of LAMaS to the main hyperparameters that control latency pressure, graph structure, and the accuracy floor.

\subsection{Parameter Sensitivity Analysis}

We analyze the sensitivity of LAMaS to four key hyperparameters: the latency penalty weight $\lambda_t$, the activation threshold $\tau$, the number of supernet layers $L$, and the accuracy floor $S_0$. We report results on HumanEval and MMLU-Pro, with the default configuration ($\lambda_t{=}0.005$, $\tau{=}0.3$, $L{=}4$) marked by $\blacklozenge$.

\begin{table}[h]
\centering
\small
\caption{Sensitivity to the latency penalty weight $\lambda_t$. Increasing $\lambda_t$ consistently reduces latency, with mild accuracy variation across the sweep.}
\label{tab:sens_lambda}
\begin{tabular}{clcc}
\toprule
Dataset & $\lambda_t$ & Acc (\%) $\uparrow$ & Latency (s) $\downarrow$ \\
\midrule
\multirow{5}{*}{HumanEval}
& 0 & 95.67 & 35.62 \\
& 0.001 & \textbf{95.93} & 23.50 \\
& 0.005\,$\blacklozenge$ & 95.42 & 21.00 \\
& 0.01 & 94.91 & 13.91 \\
& 0.05 & 93.89 & \textbf{11.29} \\
\midrule
\multirow{5}{*}{MMLU-Pro}
& 0 & 66.33 & 22.46 \\
& 0.001 & \textbf{67.13} & 16.41 \\
& 0.005\,$\blacklozenge$ & 66.27 & 14.81 \\
& 0.01 & 66.13 & 14.11 \\
& 0.05 & 65.27 & \textbf{11.38} \\
\bottomrule
\end{tabular}
\end{table}

\begin{table}[h]
\centering
\small
\caption{Sensitivity to the activation threshold $\tau$. A higher threshold activates more operators per step, increasing accuracy and latency. A lower threshold produces leaner graphs with lower latency but reduced accuracy.}
\label{tab:sens_threshold}
\begin{tabular}{clcc}
\toprule
Dataset & $\tau$ & Acc (\%) $\uparrow$ & Latency (s) $\downarrow$ \\
\midrule
\multirow{4}{*}{HumanEval}
& 0.1 & 93.89 & \textbf{13.57} \\
& 0.2 & 94.91 & 15.35 \\
& 0.3\,$\blacklozenge$ & 95.42 & 21.00 \\
& 0.4 & \textbf{95.67} & 22.80 \\
\midrule
\multirow{4}{*}{MMLU-Pro}
& 0.1 & 65.13 & \textbf{13.42} \\
& 0.2 & 65.93 & 14.23 \\
& 0.3\,$\blacklozenge$ & 66.27 & 14.81 \\
& 0.4 & \textbf{66.67} & 15.62 \\
\bottomrule
\end{tabular}
\end{table}

\begin{table}[h]
\centering
\small
\caption{Sensitivity to the number of supernet layers $L$. More layers provide greater reasoning depth and slightly higher accuracy, but increase latency due to deeper critical paths.}
\label{tab:sens_layers}
\begin{tabular}{clcc}
\toprule
Dataset & $L$ & Acc (\%) $\uparrow$ & Latency (s) $\downarrow$ \\
\midrule
\multirow{4}{*}{HumanEval}
& 2 & 94.15 & \textbf{12.85} \\
& 4\,$\blacklozenge$ & 95.42 & 21.00 \\
& 6 & 95.67 & 23.50 \\
& 8 & \textbf{95.93} & 25.54 \\
\midrule
\multirow{4}{*}{MMLU-Pro}
& 2 & 64.93 & \textbf{12.81} \\
& 4\,$\blacklozenge$ & 66.27 & 14.81 \\
& 6 & 66.47 & 16.56 \\
& 8 & \textbf{66.53} & 17.43 \\
\bottomrule
\end{tabular}
\end{table}

\paragraph{Latency weight $\lambda_t$.}
Table~\ref{tab:sens_lambda} shows that increasing $\lambda_t$ monotonically reduces latency, while accuracy varies only mildly across the sweep. On HumanEval, raising $\lambda_t$ from 0 to 0.05 reduces latency from 35.62s to 11.29s with at most 2.0\% accuracy change; the trend is consistent on MMLU-Pro. The $\lambda_t{=}0$ setting (no latency objective during training) leaves accuracy near the LAMaS level but yields the largest latency, complementing the corresponding row in the ablation study (Table~\ref{tab:ablation}). Overall, $\lambda_t$ provides smooth and predictable control over latency while keeping accuracy broadly stable.

\paragraph{Activation threshold $\tau$.}
Table~\ref{tab:sens_threshold} reveals that the threshold controls the width of the execution graph. A lower threshold ($\tau{=}0.1$) produces lean graphs with fewer operators per step, yielding the lowest latency but reduced accuracy. A higher threshold ($\tau{=}0.4$) activates more operators, improving accuracy through greater exploration together with increased latency. The default $\tau{=}0.3$ provides a balanced setting across both benchmarks.

\paragraph{Number of layers $L$.}
Table~\ref{tab:sens_layers} shows that increasing the number of supernet layers provides greater reasoning depth. On HumanEval, accuracy improves from 94.15\% ($L{=}2$) to 95.93\% ($L{=}8$), but latency increases from 12.85s to 25.54s due to the deeper critical path. The marginal accuracy gain diminishes beyond $L{=}4$, while latency continues to grow, justifying the default choice of $L{=}4$.

\paragraph{Accuracy floor $S_0$.}
Unlike $\lambda_t$, $\tau$, and $L$, which control the latency penalty and graph structure, the accuracy floor $S_0$ sets the target accuracy level used during training.
Table~\ref{tab:sens_s0} reports results across $S_0 \in \{0.95, 1.00, 1.02, 1.05\} \cdot S_{\text{ref}}$, where $S_{\text{ref}}$ denotes the orchestrator's reference accuracy and the default deployment ($S_0 = S_{\text{ref}}$) is marked by $\blacklozenge$.
The trend is monotonic on both benchmarks: on HumanEval, sweeping $S_0$ from $0.95\cdot S_{\text{ref}}$ to $1.05\cdot S_{\text{ref}}$ moves accuracy from 94.66\% to 96.18\% while latency rises from 17.85s to 23.54s; the same direction holds on MMLU-Pro (65.40\%$\to$67.13\%, 12.60s$\to$16.51s).

\begin{table}[h]
\centering
\small
\caption{Sensitivity to the accuracy floor $S_0$. We vary $S_0$ as a fraction of the architecture's reference accuracy $S_{\text{ref}}$ ($S_0 = S_{\text{ref}}$ is the default, marked by $\blacklozenge$).}
\label{tab:sens_s0}
\begin{tabular}{clcc}
\toprule
Dataset & $S_0 / S_{\text{ref}}$ & Acc (\%) $\uparrow$ & Latency (s) $\downarrow$ \\
\midrule
\multirow{4}{*}{HumanEval}
& 0.95 & 94.66 & \textbf{17.85} \\
& 1.00\,$\blacklozenge$ & 95.42 & 21.00 \\
& 1.02 & 95.67 & 22.00 \\
& 1.05 & \textbf{96.18} & 23.54 \\
\midrule
\multirow{4}{*}{MMLU-Pro}
& 0.95 & 65.40 & \textbf{12.60} \\
& 1.00\,$\blacklozenge$ & 66.27 & 14.81 \\
& 1.02 & 66.67 & 15.46 \\
& 1.05 & \textbf{67.13} & 16.51 \\
\bottomrule
\end{tabular}
\end{table}

\section{Generalizability Across Backbones and Serving Setups}
\label{app:qwen}

To assess whether LAMaS's latency advantage transfers across different LLM backbones and serving setups, we replicate the comparison on two additional configurations: the closed-source flash variant \texttt{qwen3.5-flash} served through OpenRouter, and the open-weight \texttt{qwen3.5-9b} served locally with vLLM. Both are evaluated on HumanEval and MMLU-Pro against the strongest learning-based MAS baseline (MaAS) and LAMaS. As shown in Table~\ref{tab:qwen}, LAMaS reduces latency by about 60\% on HumanEval and about 43\% on MMLU-Pro across the two settings, while matching or slightly exceeding MaAS in accuracy. This confirms that LAMaS's mechanisms transfer across distinct backbones and serving setups.

\begin{table}[h]
\centering
\small
\caption{Generalizability across Qwen backbones and serving setups: closed-source \texttt{qwen3.5-flash} served through OpenRouter and open-weight \texttt{qwen3.5-9b} served locally with vLLM. Latency is end-to-end wall-clock time per query (s). Best mean among MAS methods per column per dataset is in \textbf{bold}.}
\label{tab:qwen}
\begin{tabular}{lllcc}
\toprule
Dataset & Backbone & Method & Acc (\%) $\uparrow$ & Latency (s) $\downarrow$ \\
\midrule
\multirow{4}{*}{HumanEval}
& \multirow{2}{*}{\texttt{qwen3.5-flash}} & MaAS & 94.91 & 100.87 \\
& & \textbf{LAMaS} & \textbf{96.95} & \textbf{38.93} \\
\cmidrule{2-5}
& \multirow{2}{*}{\texttt{qwen3.5-9b}} & MaAS & 95.17 & 105.18 \\
& & \textbf{LAMaS} & \textbf{95.93} & \textbf{42.18} \\
\midrule
\multirow{4}{*}{MMLU-Pro}
& \multirow{2}{*}{\texttt{qwen3.5-flash}} & MaAS & 82.40 & 174.60 \\
& & \textbf{LAMaS} & \textbf{82.93} & \textbf{98.04} \\
\cmidrule{2-5}
& \multirow{2}{*}{\texttt{qwen3.5-9b}} & MaAS & 71.20 & 178.62 \\
& & \textbf{LAMaS} & \textbf{72.93} & \textbf{102.07} \\
\bottomrule
\end{tabular}
\end{table}

\section{Additional Baseline Comparisons}
\label{app:additional_baselines}

This section reports two additional comparisons: single-agent agentic baselines, and naive latency-cutting rules applied to an existing MAS baseline.

\subsection{Single-Agent Agentic Baselines}
\label{app:single_agent}

The non-MAS baselines in Table~\ref{tab:baseline_comparison} are prompting and sampling methods. We additionally evaluate three single-agent agentic baselines, equipping each task with the tool or self-correction mechanism it naturally supports. \textbf{Tool-augmented CoT} on GSM8K writes and executes code and answers from the execution result (the \textbf{Programmer} operator). \textbf{Test-time repair} on HumanEval runs the generated program and rewrites it from the failure feedback for up to three iterations (the \textbf{Test} operator). \textbf{Self-Refine} on MMLU-Pro critiques and revises its own answer in one round (the \textbf{SelfRefine} operator). All three reuse the operator implementations and prompts of Appendix~\ref{app:operator_set} and are evaluated under the protocol of Section~4.1.

\begin{table}[h]
\centering
\small
\caption{Comparison with single-agent agentic baselines. LAMaS rows are taken from Table~\ref{tab:baseline_comparison}. Cost is in units of $\times 10^{-4}$ USD; latency is end-to-end wall-clock time per query (in seconds).}
\label{tab:single_agent}
\begin{tabular}{llccc}
\toprule
Dataset & Method & Acc (\%) $\uparrow$ & Cost ($\times 10^{-4}$) $\downarrow$ & Latency (s) $\downarrow$ \\
\midrule
\multirow{2}{*}{GSM8K}
& Tool-augmented CoT & 89.79 & 11.38 & 18.27 \\
& LAMaS & 93.65 & 7.71 & 11.73 \\
\midrule
\multirow{2}{*}{HumanEval}
& Test-time repair & 86.51 & 6.35 & 15.43 \\
& LAMaS & 95.42 & 12.92 & 21.00 \\
\midrule
\multirow{2}{*}{MMLU-Pro}
& Self-Refine & 57.13 & 5.51 & 15.19 \\
& LAMaS & 66.27 & 10.93 & 14.81 \\
\bottomrule
\end{tabular}
\end{table}

As shown in Table~\ref{tab:single_agent}, LAMaS is 3.86~pp more accurate than Tool-augmented CoT on GSM8K, 8.91~pp more accurate than Test-time repair on HumanEval, and 9.14~pp more accurate than Self-Refine on MMLU-Pro; on GSM8K it is also cheaper and faster. These baselines and SC (Table~\ref{tab:baseline_comparison}) provide complementary controls: the agentic baselines test tool use and self-correction, and SC tests additional sampling and ensembling. SC incurs higher cost than LAMaS on all four benchmarks yet remains 9.16 and 7.60~pp less accurate on HumanEval and MMLU-Pro, respectively. Since the operator implementations are shared, what differs between these baselines and LAMaS is which operators run and how they are arranged.

\subsection{Naive Latency-Cutting Rules on an Existing MAS Baseline}
\label{app:naive_rules}

The ablation in Table~\ref{tab:ablation} covers naive latency optimization at training time (\emph{naive lat. only}). To examine how an existing MAS baseline behaves under simple latency-cutting rules at inference time, we evaluate GDesigner under two rules that require no training. \textbf{Consensus-2 stop} terminates all not-yet-launched agents as soon as any two completed agents produce agreeing answers under the agreement rule of Section~\ref{sec:gate}. \textbf{Half-agents} randomly removes $\lfloor n/2 \rfloor$ of the $n$ agents while retaining the final aggregation step. Both rules are evaluated on GSM8K and MMLU-Pro under the protocol of Section~4.1.

\begin{table}[h]
\centering
\small
\caption{GDesigner under two naive inference-time latency-cutting rules. The GDesigner and GDesigner+LAMaS rows are taken from Tables~\ref{tab:baseline_comparison} and~\ref{tab:generalizability}. Cost is in units of $\times 10^{-4}$ USD; latency is end-to-end wall-clock time per query (in seconds).}
\label{tab:naive_rules}
\begin{tabular}{llccc}
\toprule
Dataset & Method & Acc (\%) $\uparrow$ & Cost ($\times 10^{-4}$) $\downarrow$ & Latency (s) $\downarrow$ \\
\midrule
\multirow{4}{*}{GSM8K}
& GDesigner & 93.30 & 17.86 & 32.31 \\
& GDesigner + consensus-2 stop & 92.61 & 13.97 & 26.92 \\
& GDesigner + half-agents & 91.37 & 11.24 & 21.37 \\
& GDesigner+LAMaS & 93.27 & 12.08 & 20.55 \\
\midrule
\multirow{4}{*}{MMLU-Pro}
& GDesigner & 64.27 & 25.03 & 46.66 \\
& GDesigner + consensus-2 stop & 62.93 & 19.02 & 39.48 \\
& GDesigner + half-agents & 61.13 & 15.48 & 29.34 \\
& GDesigner+LAMaS & 64.33 & 17.03 & 28.27 \\
\bottomrule
\end{tabular}
\end{table}

Table~\ref{tab:naive_rules} shows that neither rule matches GDesigner+LAMaS. Consensus-2 stop is dominated on accuracy, cost, and latency, while half-agents trades accuracy for cost: at nearly the same latency, it is cheaper but 1.90 and 3.20~pp less accurate on GSM8K and MMLU-Pro, respectively. Naive latency cutting, whether at training time (Table~\ref{tab:ablation}) or at inference time (Table~\ref{tab:naive_rules}), sacrifices accuracy; among these variants, only the LAMaS pipeline reduces latency while preserving it.

\ifdefined\ARXIV\else
  \clearpage
  \input{checklist.tex}
\fi

\end{document}